\documentclass[a4paper,fleqn,preprint]{cas-dc}

\usepackage[numbers,sort&compress]{natbib}

\def\tsc#1{\csdef{#1}{\textsc{\lowercase{#1}}\xspace}}
\tsc{WGM}
\tsc{QE}

\usepackage[english]{babel}
\usepackage[utf8]{inputenc}
\usepackage[T1]{fontenc}
\usepackage{tgtermes}
\usepackage{cleveref}
\crefname{subsection}{subsection}{subsections}
\usepackage{todonotes}
\usepackage{placeins}

\usepackage{multirow}
\usepackage{siunitx}
\usepackage{graphicx}
\usepackage{dcolumn}

\DeclareSIUnit\angstrom{\text {Å}}

\usepackage[caption=false]{subfig}
\usepackage{bm}

\DeclareSIUnit{\atmospheric}{atm}

\begin{document}
\let\WriteBookmarks\relax
\def\floatpagepagefraction{1}
\def\textpagefraction{.001}

\shorttitle{Criticality in an imidazolium ionic liquid fully wetting a sapphire support}    

\shortauthors{H\"ollring et al.}  

\title [mode = title]{Criticality in an imidazolium ionic liquid fully wetting a sapphire support }

\author[1]{Kevin H\"ollring}[orcid=0000-0002-9497-3254]
\credit{Conceptualization, Methodology, Software, Formal analysis, Investigation, Writing - Original Draft, Visualization, Data Curation}


\author[2,1]{Nataša Vučemilović-Alagić}[orcid=0000-0001-5841-7181]
\credit{Simulations, Formal analysis, Visualization}

\author[2]{David M. Smith}[orcid=0000-0002-5578-2551]
\credit{Design of MD simulations}

\author[1,2]{Ana-Sunčana Smith}[orcid=0000-0002-0835-0086]
\credit{Conceptualization, Investigation, Writing - Review \& Editing, Project coordination, Funding acquisition, Resources, Data Curation}
\ead{ana-suncana.smith@fau.de, asmith@irb.hr}

\cormark[1]
\fnmark[1]

\affiliation[1]{organization={PULS Group, Institute for Theoretical Physics, FAU Erlangen-N\"urnberg},
addressline={Cauerstra\ss{}e 3},
postcode={91058}, 
city={Erlangen},
country={Germany}}

\affiliation[2]{organization={Group of Computational Life Sciences, Department of Physical Chemistry, Ru\dj{}er Bo\v{s}kovi\'{c} Institute},
addressline={Bijeni\v{c}ka 54},
city={Zagreb},
postcode={10000},
country={Croatia}}

\cortext[1]{Ana-Sunčana Smith}
\fntext[1]{Tel: +49 91318570565; Fax: +49 91318520860}

\date{\today}

\begin{abstract}
\indent
\paragraph{Hypothesis:}
Ionic liquids have various applications in catalytic reaction environments. 
In those systems, their interaction with interfaces is key to their performance as a liquid phase. 
We hypothesize that the way a monolayer ionic liquid phase interacts with interfaces like a sapphire substrate is significantly dependent on temperature and that critical behavior can be observed in the structural properties of the liquid film.
\paragraph{Methods and simulations:}
We perform molecular dynamics simulations of imidazolium-based ionic liquid monolayers deposited on a sapphire substrate at temperatures from \SI{200}{\kelvin} to \SI{400}{\kelvin}. 
We develop computational tools to analyze structural properties of molecular arrangement in the monolayer, the structure of the film and the defects spontaneously forming and healing.
\paragraph{Findings:}
We observe a clear structural phase transition at around \SI{300}{\kelvin} from a solid-like to a liquid-like behavior of a film. 
Below the critical point an alternating crystalline structure of cations and anions with alignment of periodic vectors with the underlying substrate grid is observed, with frozen defects. 
Above the critical temperature, the pattern becomes isotropic within the contact layer that displays dynamic defects of a characteristic size.
Our results highlight the importance of confinement to the phase behavior of the system. 
\end{abstract}

\begin{graphicalabstract}
\includegraphics{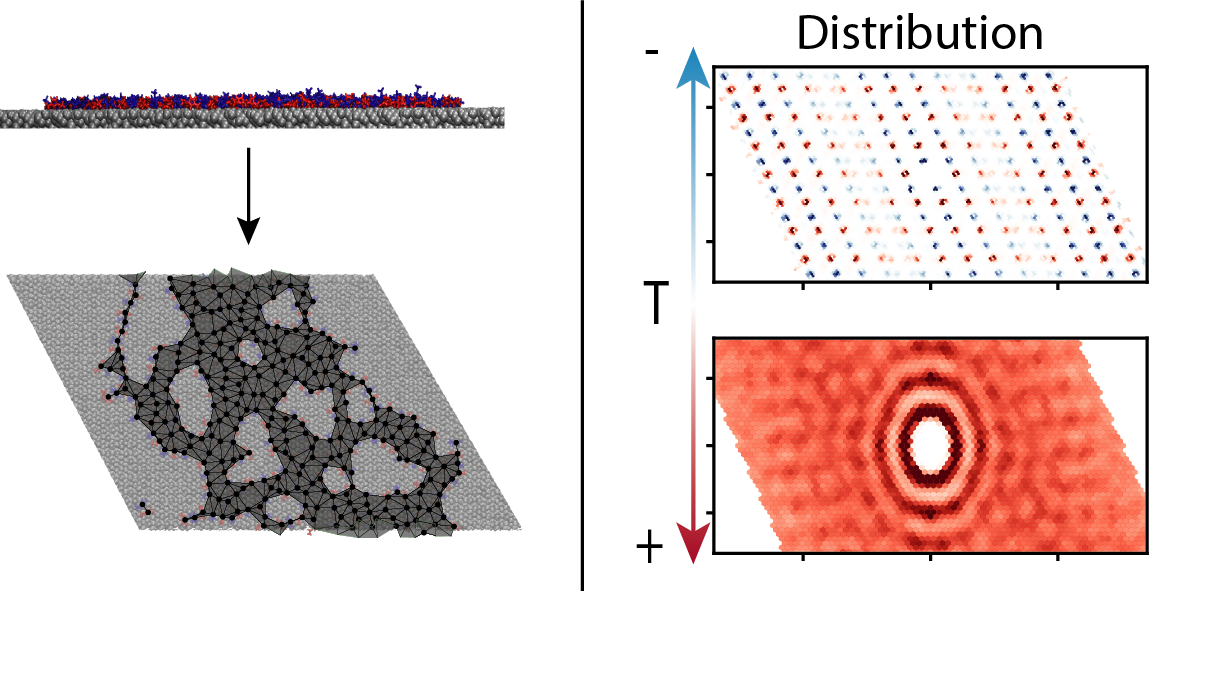}
\end{graphicalabstract}



\maketitle

\section{Introduction}
Ionic liquids (ILs) are molten salts whose liquid state spans a wide temperature range including room temperature \cite{freemantle2010introduction}. 
The charged nature of their ionic molecular constituents makes Coulomb interactions a major influence on their behavior and structure. 
Their liquid state at room temperature is a consequence of these Coulomb interactions competing with geometrical constraints preventing the formation of a regular and rigid crystal grid due to the sizes of the involved ions. 
Based on these chemical and structural properties, they have wide applications in various industrial scenarios \cite{lexow2020ultrathin}. 
In addition to their use as lubricants \cite{cai2020ionic,somers2013review,zhou2009ionic} and in chemical reaction engineering \cite{freemantle2010introduction,hao2007self,hubbard2011understanding,plechkova2008applications}, they have also been employed as an electrolyte \cite{francis2020lithium,sun2019safe}. 

For most applications, in addition to the liquid state and the highly polar nature of ILs, the presence of a solid supporting the liquid phase is of vital importance.
Specific examples include the surfaces of catalysts involved in chemical reactions such as the supported ionic liquid phase (SILP) \cite{riisager2006supported} or the solid catalyst with ionic liquid layer (SCILL) systems \cite{misra2020water}. 
The importance of the support, whether functional or not, emerges due to the strong interfacial structuring of ILs due to adhesive forces, the intensity of which depends on the nature of the interactions between IL ions and the exposed surface layer. 
Ultimately, these interactions determine the wetting behavior of the IL \cite{vucemilovic2019insights}, the mobility of the ions and the reactants \cite{epm_paper}, and may affect reaction rates. 

On neutral surfaces like fully hydroxylated alumina, IL adsorption is mainly governed by the formation of hydrogen bonds leading to a coupling between the liquid layer structure and the underlying substrate \cite{segura2013adsorbed, vucemilovic2019insights}.
In films thicker than \SI{10}{\nano\meter}, this leads to several well defined solvation shells propagating towards the bulk of IL \cite{lhermerout2018ionic,perez2017scaling,vucemilovic2019insights,brkljaca2015complementary, horn1988double,atkin2007structure}. 
In thin films, it has been shown that the organization of the film depends on the film thickness, while the structure of the contact layer could be investigated in appreciable detail using a variety of experimental techniques  \cite{lexow2020ultrathin,steinrueck2015ionic,earle2006distillation,cremer2008physical,souda2008glass}.

For example, with regard to the wetting process of gold and platinum surfaces by [C$_1$C$_1$Im][NTf$_2$], characteristic changes in the monolayer structure could be measured depending on the degree of surface saturation \cite{meusel2020growth} as well as depending on the temperature of the system \cite{meusel2020atomic,meusel2020growth}.
At low temperatures with an unsaturated surface, the formation of islands of the IL film on the substrate was observed with overall hexagonal and occasionally striped crystalline patterns of anion distribution. 
Further analysis showed that the crystalline-like phase can also be stabilized in an IL bilayer. 
However, a precise analysis of the structural phase transition in these monolayer systems was limited by the resolution of the imaging method and the lack of information on the cation distribution. 
Nonetheless, this ensemble of work pointed out the importance of the thermodynamic conditions and the amount of IL deposited per unit surface area for the emergent patterns. 

The issue of the resolution can be resolved and the full understanding of the organization of IL ultra-thin films can be obtained in molecular dynamics (MD) simulations, yet the organization of such ultra-thin films has been studied only sparsely using computational methods \cite{maruyama2018ionic}.
In this publication, we therefore focus on MD simulations and study the effects of temperature and saturation on the organisation of IL monolayers. 
Our system of choice is [C$_2$Mim][NTf$_2$] deposited on a hydroxylated sapphire crystal, as a paradigmatic example of a setting where full wetting is expected. 
While focusing on the characterization of the structural properties of the films with a thickness of as little as one molecular layer, we show that the temperature affects the balance between cohesive and adhesive forces acting on the films' constituents, causing different morphological and periodic behavior never before unraveled at this level of detail.

\begin{figure*}
    \centering
    \includegraphics[width=1\textwidth]{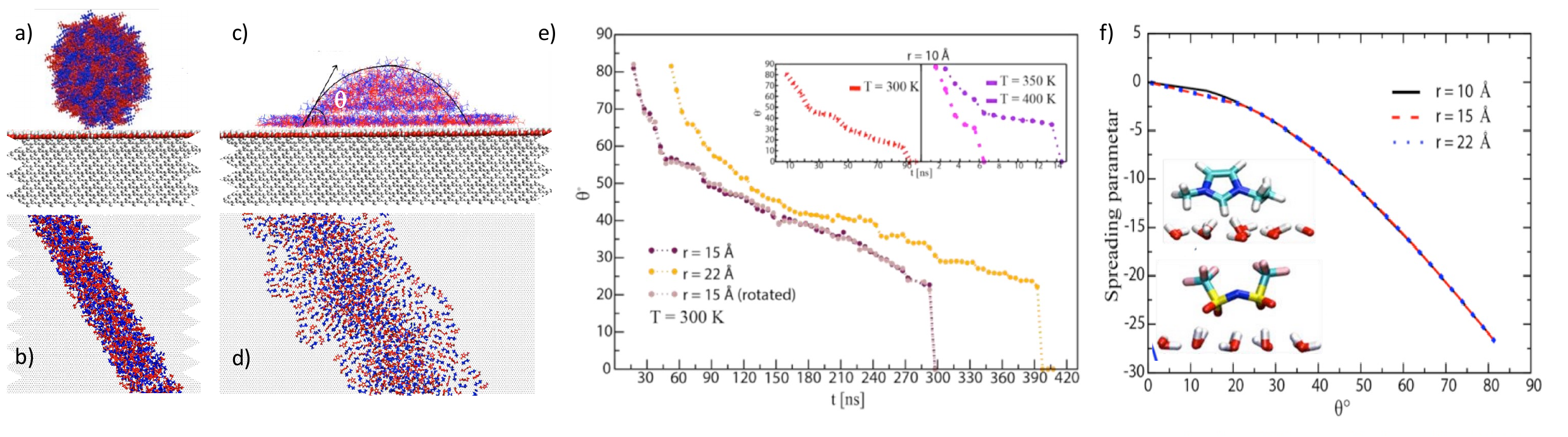}
    \caption{\textbf{Generating monolayers by complete wetting.}
    Nanodroplet brought in contact with the hydroxylated sapphire surface seen at the onset of the wetting simulations a) along the (1000) axis and b) from the top in the $xy$-plane. 
    The same perspectives are shown during spreading in c) where we also present the effective spreading angle and in d) where the structure of the precursor monolayer can be clearly seen. 
    e) The effective spreading angle as a function of time for nanodroplets of different sizes at \SI{300}{\kelvin} and at different temperatures (see insets) is shown. 
    The typical orientation of the nanodroplet is along the (1000) axis, except for the data set marked as "rotated" in the legend.
    f) Spreading parameter as a function of the effective spreading angle for nanodroplets of different sizes at $T=\SI{300}{\kelvin}$. 
    Insets show typical orientations of the cation (top) and anion (bottom) forming hydrogen bonds with hydroxyl groups on the sapphire surface. 
    (Figures adapted from the PhD thesis of an author. \cite{vuvcemilovic2021computational})
    }
    \label{fig:wetting}
\end{figure*}

\section{Simulation protocols}

The simulations of monolayers were executed in {GROMACS} 5.1.2 \cite{van2005gromacs}. 
The IL was built from [C$_2$Mim]$^+$ and [NTf$_2$]$^-$ ions for which the force field parameters and charges were taken from previous work \cite{vucemilovic2019insights,vuvcemilovic2021computational,epm_paper}.
As suggested in these publications, to recover the experimentally validated organization of the sapphire-IL interface, all atomic charges of IL ions had to be scaled by a factor of $0.9$, while the van der Waals interactions between the IL and the sapphire were calculated using the Lorentz-Berthelot rules. 
The sapphire was parameterized using the CLAYFF force field \cite{CLAYFF}, fully hydroxylated using previously developed parameters \cite{brkljaca2015complementary} and optimized with GULP \cite{GULP}. 

All simulations were performed using periodic boundary conditions with a monoclinic geometry, to respect the symmetry of the crystal.
We systematically use a time step of \SI{2}{\femto\second} and a cut-off of \SI{2}{\nano\meter} for the van der Waals and short-range Coulomb interactions, while the Particle Mesh Ewald procedure was employed for the description of the long-range Coulomb interactions. 
Periodic copies of the slabs of alumina with the IL film were systematically separated by an \SI{80}{\nano\meter} vacuum slab to eliminate long-range Coulomb forces that could affect the interfaces. 
The desired temperature was maintained with the Nose-Hoover thermostat.  

Full monolayers were created in wetting simulations. 
For those, we start with a box of a fully equilibrated bulk liquid at the desired temperature \cite{vucemilovic2019insights,vuvcemilovic2021computational,epm_paper}. 
From this box, using VMD, we extract nanodroplets that are circular in the $xz$ plane ($r = \SI{10}{\angstrom}$, $154$ ion pairs), while in the $xy$ plane the droplet is positioned to follow the $(1000)$ crystal axis ( see \cref{fig:wetting}a+b). 
After putting the droplet in close proximity of the sapphire slab (\SI{46.5462}{\nano\meter} $\times$ \SI{13.4364}{\nano\meter} $\times$ \SI{2.12}{\nano\meter} in $x$, $y$ and $z$ directions), the wetting simulation is performed in the NVT ensemble (\cref{fig:wetting}c+d), at temperatures of \num{250}, \num{300}, \num{350} and \SI{400}{\kelvin} for a typical duration of \SI{100}{\nano\second}. 
The effect of the cylinder size was tested at $T=\SI{300}{\kelvin}$, where additional nanodroplets of radius \SI{15}{\angstrom}, and \SI{22}{\angstrom} were carved and their full spreading was simulated for \num{300} and \SI{400}{\nano\second} respectively.
An additional simulation of a nanodroplet of \SI{15}{\angstrom} rotated by $30^\circ$ was performed at \SI{300}{\kelvin} to quantify the sensitivity to the initial spreading direction. 
For all droplet sizes at $T=\SI{300}{\kelvin}$, the spreading process followed a universal curve of the spreading parameter $S$, defined as $$S = \gamma(\cos(\theta)-1).$$ 
This makes $S$ a function of the time dependent, effective droplet contact angle $\theta$ (\cref{fig:wetting}e) as well as the surface tension $\gamma$. 
The evolution of $S(\theta)$ (\cref{fig:wetting}f), shows the insensitivity of the wetting process to the initial conditions.
Once the monolayers were obtained, a \SI{100}{\nano\second} run in the NVT ensemble was performed for equilibration. 
The ensuing production run for the analysis of the monolayer structure was performed for up to \SI{200}{\nano\second}.

An alternative procedure was also established to verify that the structure of the monolayers does not depend on the protocol with which they were created, and to investigate systems at temperatures at \SI{250}{\kelvin} and below, where spreading is very slow. 
Specifically, monolayers were created starting from a \SI{16}{\nano\meter} thick ionic liquid film of $1800$ pairs of [C$_2$Mim]$^+$ and [NTf$_2$]$^-$ ions deposited on a sapphire spanning  \SI{7.57}{\nano\meter} $\times$ \SI{6.29}{\nano\meter} $\times$ \SI{2.12}{\nano\meter} in $xyz$ directions \cite{epm_paper}. 
The system was constructed from a box of equilibrated IL placed on top of the sapphire, re-equilibrated to the temperature of \SI{300}{\kelvin} using semi-isotropic annealing along the $z$ axis in NPT\cite{vucemilovic2019insights}, followed by an at least \SI{200}{\nano\second} NVT run. 
After the thick film was created at \SI{300}{\kelvin}, all but the first molecular layer of the IL film was removed. 
This resulted in $54$ ion pairs on the sapphire surface which was, consequently, extended along the $x$-axis, to obtain an $xy$ dimension of $15.14\times \SI{6.29}{\nano\meter\squared}$, and thus, a low level of surface saturation. 
The monolayer was again equilibrated, and treated as described above. 

We furthermore calculate the potential of mean force (PMF) that an ion pair in the monolayer experiences. 
This was achieved by pulling away the center of mass (COM) of an ion pair (or an ion) from a fully equilibrated monolayer in a direction perpendicular to the surface of the monolayer i.e. along the $z$-axis.
Sampling is achieved by first performing a \SI{400}{\pico\second} pull during which the COM is attached to a virtual particle (spring constant of \SI{12000}{\kilo\joule\per\mol\per\square\nano\meter} ) moving at a constant velocity of \SI{0.01}{\nano\meter\per\nano\second}. 
From the obtained trajectory, we extract $14$ configurations, wherein each successive configuration features the COM separated from the surface by an additional \SI{0.02}{\nano\meter} along the $z$ axis. 
At larger distances, $10$ configurations are, furthermore, extracted for which the separation between successive COMs is increased to \SI{0.1}{\nano\meter}.
These $24$ configurations are used for umbrella sampling, with a force constant of \SI{5000}{\kilo\joule\per\mol\per\square\nano\meter} applied to configurations distant from the surface, whereas a value of \SI{10000}{\kilo\joule\per\mol\per\square\nano\meter} was employed for configurations near the surface, in total covering separations of \SI{1.85}{\nano\meter} from the positions within the monolayer. 
Each sampling configuration was propagated for \SI{10}{\nano\second}. 
Data from the first \si{\nano\second} were omitted from the analysis.
The weighted histogram analysis method (WHAM) was used to to obtain unbiased PMFs.

\section{Interactions of ions with the hydroxylated sapphire surface}

\begin{figure}
    \centering
    \includegraphics[width=.5\textwidth]{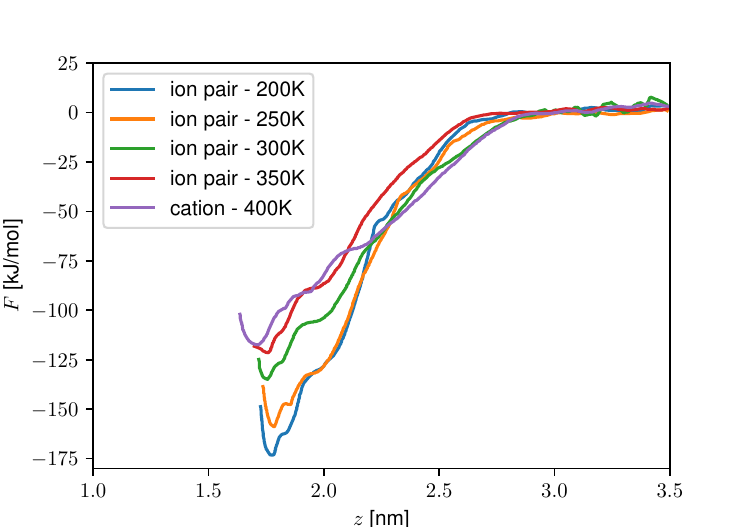}
    \caption{\textbf{Potential of Mean Force (PMF) at various temperatures.}
    The PMF of an ion pair within the monolayer is shown for temperatures from \SI{200}{\kelvin} to \SI{400}{\kelvin}.
    At \SI{400}{\kelvin}, the interaction potential is calculated exclusively for a cation. 
    The adhesion energy is observed to drop as $T$ is increased. (Figure as in the PhD thesis of an author \cite{vuvcemilovic2021computational})
    }
    \label{fig:app_pmf}
\end{figure}

We first explore the interactions of the ions with the surface. 
In agreement with previous work \cite{brkljaca2015complementary,vucemilovic2019insights}, we find that the interaction with the substrate is promoted by hydrogen bonds that are formed between the oxygen of the hydroxyl groups and the hydrogen on the C2 atom of the imidazole ring (see inset of \cref{fig:wetting}e). 
Meanwhile, the anion, predominantly in its cis-conformation, forms hydrogen bonds with its four oxygens pointing towards the surface.

During our calculations of the PMF in the temperature range $T<\SI{400}{\kelvin}$, we observe that pulling one ion type necessarily results in the extraction of an anion-cation pair, irrespective of the applied force constant or of the pull velocity. 
This is a consequence of strong cohesive Coulomb forces. 
Only at \SI{400}{\kelvin}, the cation could be pulled without accompanying anion, as shown in the Figure \cref{fig:app_pmf}.
Consequently, for the calculation of the PMF at below \SI{400}{\kelvin}, we opted to constrain the COM of an anion pair, whereas at \SI{400}{\kelvin} we only constrained the COM of a single cation.

The resulting PMFs show a deep minimum of about \SI{180}{\kilo\joule\per\mol} at \SI{200}{\kelvin} (see \cref{fig:app_pmf}) that up to \SI{400}{\kelvin} still persists at about \SI{130}{\kilo\joule\per\mol}, as a consequence of the hydrogen bonding, and the cohesion forces with other ions in the film
Such strong forces at all temperatures promote the complete wetting of the hydroxylated sapphire interface and the formation of pure monolayers. 

\begin{figure*}[b]
    \centering
    \includegraphics[width=.9\textwidth]{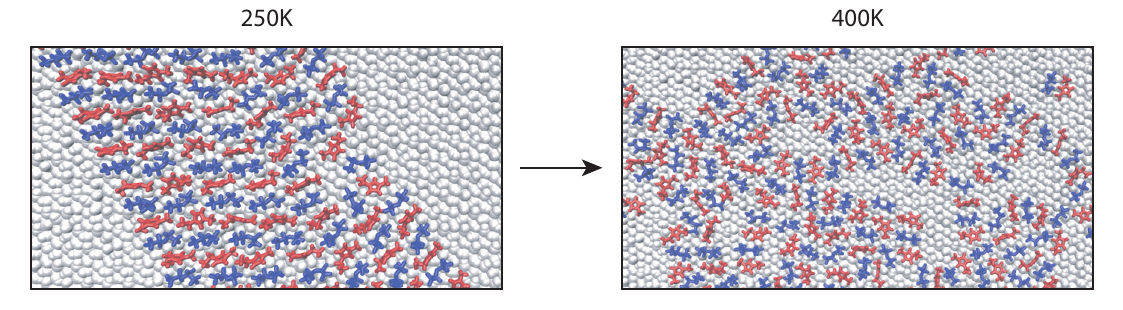}
    \caption{\textbf{Visualization of the characteristic molecular alignment and orientation below and above the critical temperature.} Cations (red) and anions (blue) on top of the hydroxylated sapphire surface (silver) at \SI{250}{\kelvin} are shown on the left and at \SI{400}{\kelvin} on the right. 
   The directional long-range order aligned with the sapphire crystal structure observed at low temperatures is lost at high temperatures, where only short-range order prevails reflected in an alternating anion/cation pattern. 
   Furthermore, the formation of defects, seen as holes in the monolayer, is promoted at higher temperatures.}
    \label{fig:transition_illustration}
\end{figure*}

\section{The effects of temperature on the organization of IL monolayers}
\begin{figure*}
    \centering
    \includegraphics[width=\textwidth]{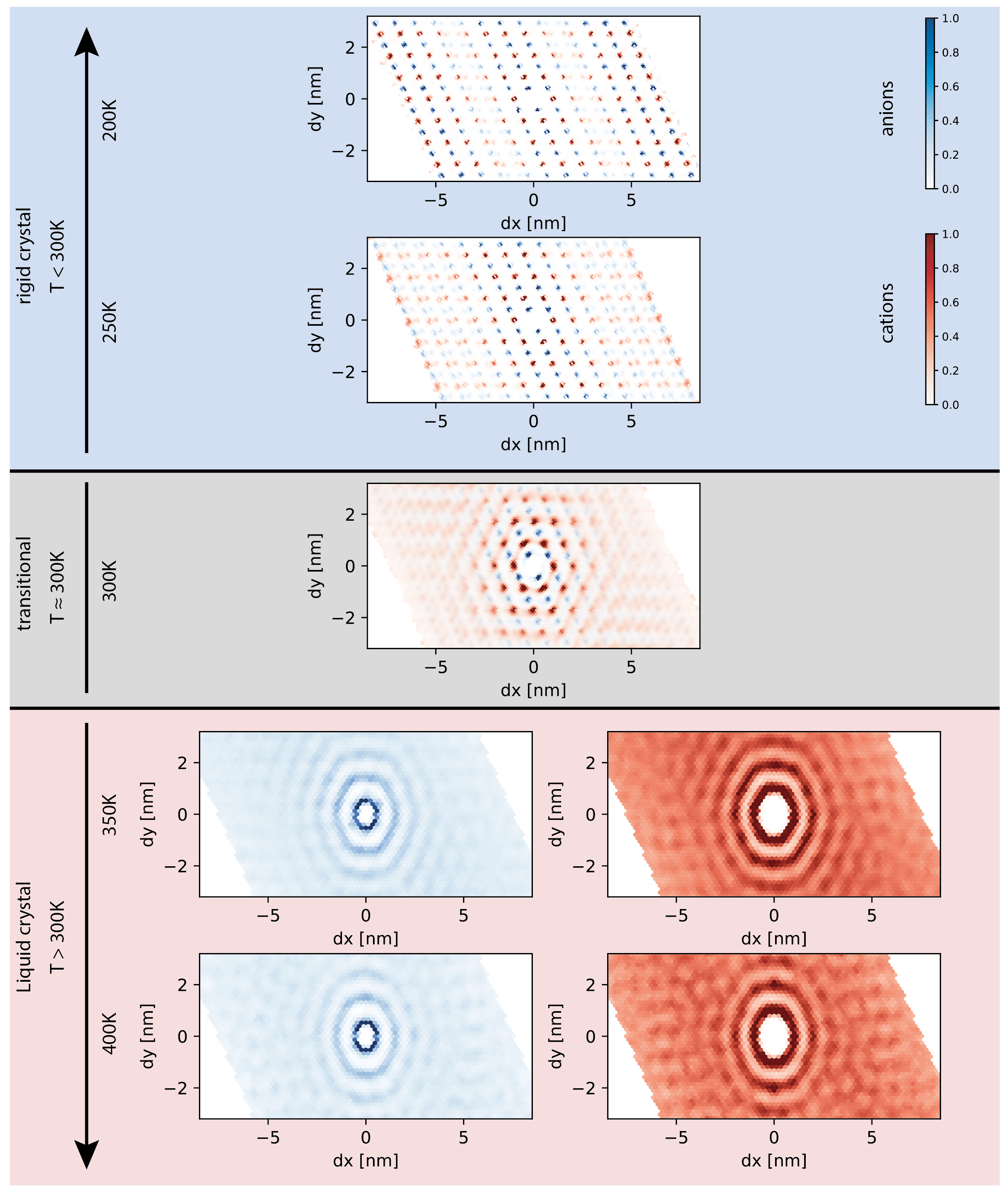}
    \caption{\textbf{Relative density distribution functions of monolayer films as a function of temperature.}
    2Drdfs with respect to a cation in the center at $dx=dy=0$ are shown for anions (blue) and cations (red). 
    Below the critical temperature threshold (\SI{200}{\kelvin} and \SI{250}{\kelvin}, top panel), we observe alternating crystal-like patterns.
    The distribution is very clearly directionally dependent. 
    The symmetry of this distribution is also closely tied to the symmetry axis of the underlying substrate. 
    The middle panel represents the structure close to the critical temperature ($\approx$\SI{300}{\kelvin}) where we observe an intermediate state between the crystal-like pattern and the direction-independent distribution at higher temperatures.
    Above the critical temperature (\SI{350}{\kelvin} to \SI{400}{\kelvin}, bottom panel), the structure is isotropic, i.e., independent of the direction in the two-dimensional plane and consequently decoupled from the structure of the substrate. 
    The number of oscillations in the density profile also decreases with higher temperatures, indicating a drop in density-density correlation.}
    \label{fig:chrystal_structure}
\end{figure*}

We now consider the spatial distribution of ion pairs in the IL on top of the solid substrate. 
From visual inspection of the trajectory alone we observe a clear alternating pattern of cations and anions aligned with the $(1000)$ axis of symmetry of the sapphire at low temperatures. 
This is contrasted by an isotropic, but still alternating pattern at higher temperatures (see \cref{fig:transition_illustration}). 

To quantify this heuristic observation, we extract the normalized relative density distribution in the $xy$-plane (2Drdf) and compare its appearance at various temperatures (see \cref{fig:chrystal_structure}). 
The 2Drdf is obtained by calculating the spatial distribution of the COM position of all cations or anions relative to an initial cation,  averaged over all possible initial cations and all frames in the production run (code available on \url{https://github.com/puls-group/monolayer_analysis} and \cite{zenodo_code_archive}). 
The distributions are binned by decomposing the positions into coefficients in the basis of the period bounding box of the system within the $x$-$y$-plane and restricting these coefficients to the interval $[-0.5, 0.5]$.
In the next step, the relative positions are binned in squares on a two-dimensional grid with $dx$ and $dy$ of $\approx \SI{0.1}{\nano\meter}$ side length. 
Because the monolayer has the appearance of an infinite strip along the $(1000)$ axis, we normalize the distribution along the $x$-axis using a rescaling factor of  $\propto 1/(a-|dx|)$, where $a$ denotes the finite width of the film in $x$-direction, which allows us to compare different axes. 
We furthermore, normalize the distributions to 1 at their maximums to allow for better visibility.

At low temperatures (\SIrange{200}{300}{\kelvin}, top panels in \cref{fig:chrystal_structure}), the relative distribution function confirms the observation of alternating cation and anion positioning along the $(1000)$ axis, with stripes of like-charged ions along the $x$-axis. 
The pattern further confirms a strong directionality of the ion distribution with two intercalated distorted hexagonal lattices, one for each ion type. 

In contrast to experiments with [C$_1$Mim][NTf$_2$] on Au(111) surfaces \cite{meusel2020atomic}, where a mismatch of the alignment between the liquid layer and the substrate has been reported, we observe long-range order appearing at temperatures below \SI{300}{\kelvin} very clearly aligned with the periodicity vectors of the underlying $(0001)$ sapphire surface, due to the distribution of hydroxyl moieties. 
In this regime, we find correlations spanning our entire IL, with nearly no movement of the ions on the surface. 
This highlights the importance of strong adhesion forces as a consequence of the formation of hydrogen bonds \cite{brkljaca2015complementary}.
The impact of hydrogen bonds on the structure of a monolayer has also recently been reported on a graphite surface, where the hydrogen bond network led to the formation of a checkerboard pattern \cite{zhang2018structure}.

At \SI{300}{\kelvin} the rigid periodic structure of the IL monolayer appears to mallow and the directional distribution is beginning to be smeared out into a more and more isotropic (i.e., radially symmetric) distribution (see \cref{fig:chrystal_structure}). 
The peaks in the distribution of both cations and anions start to connect, with strong correlations spanning only the 4 to 5 closest neighbors. 
At this point, the increased thermal excitation is able to overcome the adhesive forces fixing the ions to specific positions on the hydroxylated sapphire surface. 
This allows them to hop between the surface interaction sites more frequently, while still being adsorbed to the interface. 

At temperatures at and above \SI{350}{\kelvin} we observe a fully isotropic, i.e., direction-independent distribution of cations and anions around the reference cation (see \cref{fig:chrystal_structure}). 
The in-plane correlation length between ions decreases further at higher temperatures, as is expected. 
This can be quantified in terms of the number of visible peaks in the radial distributions dropping from 3 or 4 at \SI{350}{\kelvin} to between 2 and 3 rings being visible at \SI{400}{\kelvin}.
Notably, the circular peaks in the distributions of anions and cations respectively as visualized in \cref{fig:chrystal_structure} continue to exhibit alternating patterns between different ion species as a consequence of the Coulomb attraction between different and the repulsion between similar ion species. 
However, at the imposed density of IL ions, the adhesive forces to the sapphire still dominate, thus preventing the spontaneous formation of bi-layers, previously observed on Au(111) surfaces at only \SI{300}{\kelvin} \cite{meusel2020atomic}.

These insights show that there exists a temperature threshold, below which the structure of the IL monolayer is mostly governed by the adhesive forces between the substrate and the liquid molecules.
In contrast, above it the structure is mainly determined by IL-internal Coulomb interactions and overall film-internal cohesive forces. 
According to our results, this critical threshold for [C$_2$Mim][NTf$_2$] and the hydroxylated sapphire surface with low surface saturation is clearly situated at or around room temperature.

\section{Film morphology and defects}
\begin{figure*}
    \centering
    \includegraphics[width=.9\textwidth]{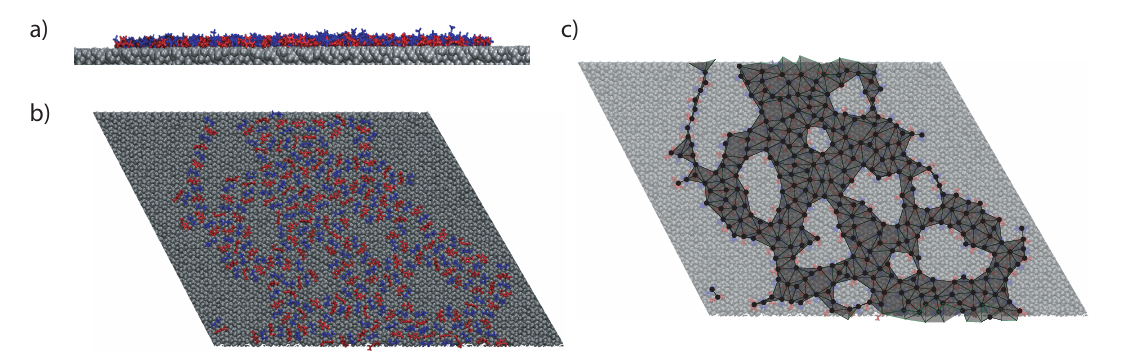}
    \caption{\textbf{Graph representation of the monolayer.}  An example from a system at $T=\SI{400}{\kelvin}$ is shown where the supported IL monolayer is shown (a) from the side and (b) from the top. 
    (c) We present the result of the segmentation of the monolayer with the detected covered area being overlaid in dark gray. }
    \label{fig:system_illustration}
\end{figure*}

We want to further quantify the impact that this structural phase transition has on the topology and morphology of the observed films as well as the defects forming therein, as the melting of the crystaline-like phase takes place. 
For this purpose, we analyze the precise shape and structure of the liquid layer within each frame.
In short, we first build a two-dimensional planar graph of the IL molecules in a periodic geometry, where we connect pairs of molecules that find themselves within a threshold distance using custom-made code provided at \url{https://github.com/puls-group/monolayer_analysis} and in the ZENODO Archive \cite{zenodo_code_archive}. 
Then, we identify the outside edge of the film and the inner boundaries of defects in the film layer (see \cref{fig:system_illustration}c for a visualization of a graph representation). 
We then analyze the properties of the resulting constructed film and its defects, as detailed in the following sections.

\subsection{Establishing the analysis procedure}
\subsubsection{Graph construction from MD data} \label{app:graph_construction}
To construct a graph from a snapshot of an IL film, each center of mass (COM) of a molecule was represented by a vertex and an edge was drawn between two vertices if their distance was below a threshold $r_\mathrm{T}$. 
During construction, COMs of molecules are added in the order of their appearance in the trajectory file. 
For each newly added entry, all pre-existing COMs with a distance less than $r_\mathrm{T}$ are determined and an attempt to insert a direct connecting edge is made. 
If that edge crosses an existing edge, the attempt is aborted to keep the graph planar. 
Once the vertices are all part of the graph, the faces of the graph are detected by following the faces surrounding the edge in a clockwise-oriented fashion. 
In doing so, inner faces of the film are detected in the correct (clockwise) orientation, whereas outer faces of the graph are found in the opposite (counter-clockwise) orientation or through the necessity of periodic copies of vertices to close their outline. 

Based on our analysis of the 2Drfs of the IL molecules at various temperatures (see \cref{fig:chrystal_structure}), we chose $r_\mathrm{T}=\SI{1.3}{\nano\meter}$ as the threshold distance for connecting vertices. 
This captures only the first two layers of the surrounding ions in all systems. 
The choice of $r_\mathrm{T}$ to be further than just the very first alternate charge layer's distance around a molecule guarantees that even without all connections being drawn, the graph of a film is correctly constructed, including the detection of holes within the film. 
Furthermore, this procedure allows us to find chains of molecules on the surface and to discern between disconnected droplets of IL which form occasionally at higher temperatures. 

\subsubsection{Morphological measures} \label{app:morphology_statistics}

With the graph of the film available, defects and holes within the film can be detected based on the number of vertices on the outline of an inner face. 
Following visual inspection, if at least $V_\mathrm{hole}=5$ vertices are found, the face is considered to be a hole. 
The area of the film ($A$) Is calculated as the sum of the area of all inner faces with less than the threshold $V_\mathrm{hole}$ of vertices. 
Similarly, the film perimeter ($P$) is calculated based on the sum of the perimeters of all outside faces as well as those of all inner faces satisfying the above hole-detection criterion. 
Alternatively, an approach with an area-based threshold criterion may also be applicable, yet the current approach is found to be computationally very effective. 

\begin{figure*}
    \centering
    \includegraphics[width=.5\textwidth]{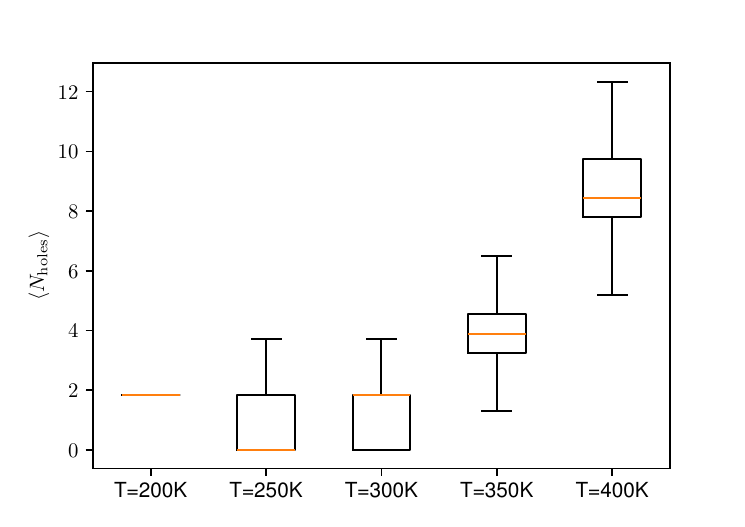}%
    \includegraphics[width=.5\textwidth]{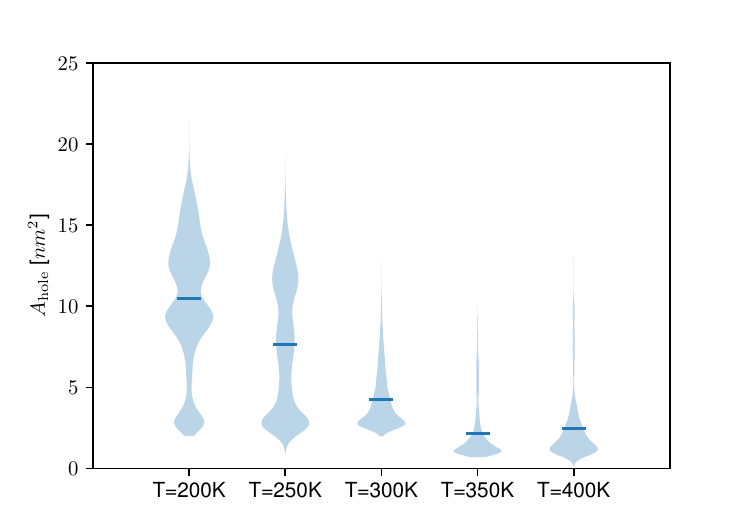}%
     \newline
    \includegraphics[width=.5\textwidth]{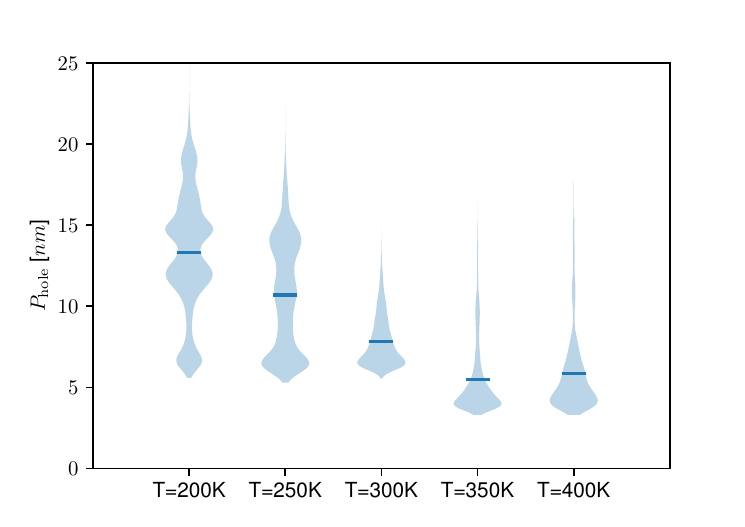}%
    \includegraphics[width=.5\textwidth]{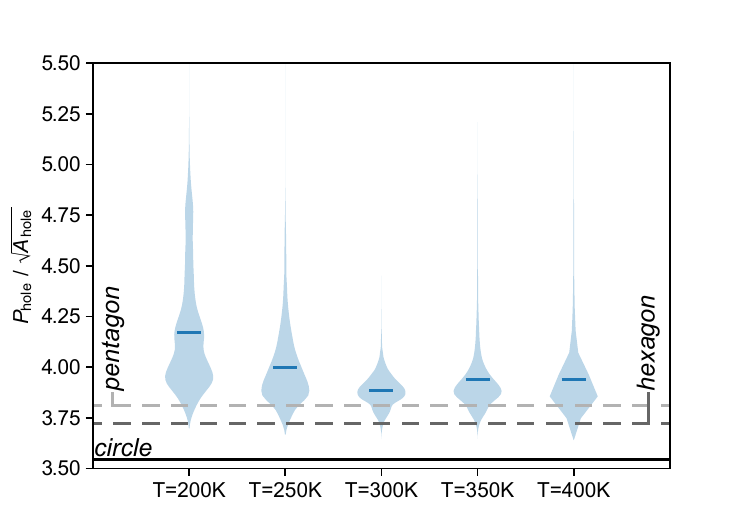}
    \caption{\textbf{Morphological analysis of defects.} 
    Defects seen as holes in the monolayer films have been identified and characterized in a statistically sound manner using custom-made code provided at \url{https://github.com/puls-group/monolayer_analysis} and in the ZENODO Archive \cite{zenodo_code_archive}. 
    (top left panel) Number of holes as a function of temperature in a box-plot. 
    The mean is represented by the orange line. 
    We see a clear transition in the number of holes with more holes appearing in the liquid state as the temperature rises in line with the criticality observed in \cref{fig:chrystal_structure}.
     Violin plot representations of the distribution of defect area (top right) and perimeter (bottom left) in monolayer films at various simulation temperatures.  
     (bottom right) The distribution of IPR values of holes in monolayers simulated at between \SI{200}{\kelvin} and \SI{400}{\kelvin} are plotted together with the theoretically minimal IPR values for circles as well as hexagon and pentagon shapes. 
    The raw area and perimeter statistics for holes are available on ZENODO \cite{zenodo_data_archive}. 
    }
    \label{fig:hole_morphology}
\end{figure*}

We finally calculate the isoperimetric ratio (IPR) $P/\sqrt{A}$ of the film as a measure how long the outline of a shape is relative to its area.
Lower values indicate a stronger tendency to contract to a circular droplet, whereas higher values indicate lower line tension and a tendency to form fingers and detached droplets. 
In non-periodic planar geometry, the IPR attains its smallest possible value $2\sqrt{\pi}$ for a circle, which optimizes the area enclosed by a fixed perimeter length.
Polygons with finite vertex counts necessarily have larger values than circles, but polygons with more vertices are generally able to better approximate a circle's IPR (see \cref{fig:hole_morphology} for limits of hexagons and pentagons). 
In periodic structures like our systems, the IPR can have even smaller values than for a circle.
On the other end of the spectrum, larger IPR values are indicative of larger holes and more rugged outlines.
We therefore calculate the number of holes $N_\mathrm{holes}$ (i.e. the number of inner faces with at least $V_\mathrm{hole}$ vertices) in each frame. 

To determine the uncertainty of our results, we calculate the mean and the standard deviation $\sigma$ of each of these properties ($\langle A\rangle$, $\langle P\rangle$, $\langle IPR\rangle$) across all frames within the trajectory. 
Finally, we also calculate the standard error $\sigma_\mathrm{err}=\sigma/\sqrt{f}$, where $f$ is the number of contributing frames.

\begin{figure}[b]
    \centering
    \includegraphics[width=.5\textwidth]{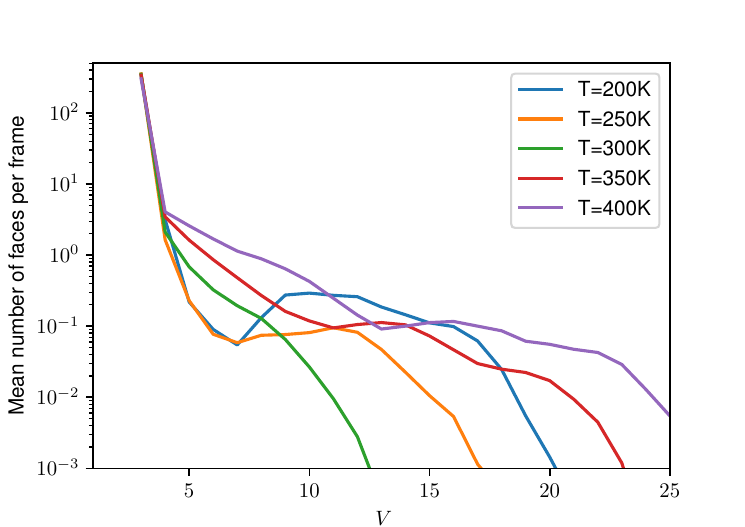}%
    \caption{\textbf{Statistics of number of vertices $V$ per inner face and holes.}
    Faces considered part of the film covering the substrate ($V<5$) are about $10^2$ times more frequent than the faces considered to be holes ($V\geq 5$).}
    \label{fig:film_hole_distribution}
\end{figure}

\subsubsection{Statistical normalization} \label{app:normalization}
Due to the different number of ion pairs at lower and higher temperatures, we present normalized statistical data.
The statistical data is specifically normalized to a virtual system with $100$ ion pairs. 
For that purpose, we assume that ideally each ion pair should cover approximately the same surface area in each system, leaving us with a scaling factor $$R=\frac{100}{N_\mathrm{pairs}}$$
for areas. 
We also use this scaling factor for the number of holes but not for the number of vertices within these holes. 
For the purpose of lengths like $P$, we use the corresponding scaling factor $\sqrt{R}$ to harmonize the results.

\subsection{Distribution of faces} \label{sec:faces}
We first calculate the distribution of faces with a particular number of vertices, averaged over all film realizations at the given temperature \cref{fig:film_hole_distribution}. 
Faces are considered to be part of the film if they have $4$ or less vertices. 
In agreement with the generally hexagonal structure, faces with 3 vertices are the most likely configuration in all systems. 
Faces with a larger number of vertices belong to the positions in the film where a defect appears, with the number of vertices being  correlated to the size of the defect, seen as a hole in the film.

At low temperatures (\SI{200}{\kelvin}, \SI{250}{\kelvin}), the highly ordered phase typically contains defects that are frozen in the film, with very little to no restructuring on the time scales of simulations. 
As the system is building alternating arrays of ions, that quasi percolate infinitely along the $(1000)$ axis, the defects appear when the number of ions is incommensurate with completing a full stripe, and are hence, somewhat a consequence of the very setup. 
At \SI{250}{\kelvin} such a defect can travel to the edge of the otherwise unperturbed monolayer still being tightly coupled to the alumina lattice. 
As the transition temperature is approached, the film can reduce the line tension, hence narrowing down the distribution of faces in all measures. 
Above the transition temperature at around \SI{300}{\kelvin}, due to the unsaturated nature of the systems, defects dynamically form and reorganize as the line tension is no longer able to sustain in-plane fluctuations of the edge of the film. 
The consequence is a broadening of the distribution of the number of faces and the appearance of faces with very large number of vertices along with a re-widening of the distribution of the morphological measures of the holes. 

This analysis shows that the temperature evolution of these distributions as visualized in \cref{fig:film_hole_distribution} clearly represents the change in the phase behavior of the IL from a solid-like crystalline to a fully liquid state. 

\subsection{Morphology of the defects} \label{sec:holes}
It can be noted that the total number of holes is relatively small compared to the number of faces that are considered to be a part of the IL layer covering the aluminum crystal. 
Specifically, we observe several hundreds of faces as a part of the monolayer in contrast to only single- or low two-digit average numbers of holes being observed per frame across all temperatures (see \cref{fig:hole_morphology}). 
The statistics, furthermore, demonstrate that there is a clear trend of the number of holes (or at least larger holes by area) within the film decreasing as the temperature approaches the transition temperature near \SI{300}{\kelvin}. 
To be more precise, at \SI{200}{\kelvin} we observe larger holes than at \SI{250}{\kelvin}, where we in turn observe more large holes than at \SI{300}{\kelvin}. 
At \SI{300}{\kelvin} and above, we see a general trend of the number of holes rising with increasing temperature.
Additionally, we also observe a temperature-dependent increase in the maximum of the largest areas and perimeters of holes being observed at these temperatures.

Focusing on the morphology of the holes, we observe a trend of shape optimization towards the critical temperature. 
Specifically, as illustrated in \cref{fig:hole_morphology}, the distribution of the isoperimetric ratio (IPR) of holes $P_\mathrm{hole}/\sqrt{A_\mathrm{hole}}$ is closest to circular shapes at $T=\SI{300}{\kelvin}$. 
Towards higher but much more notably towards lower $T$, holes with more elongated shapes can be observed. 
At these low temperatures, the shape of the holes is imposed by the tendency of the ions to follow the underlying lattice, and hence more elongated defects are created, which are very slow to relax, due to the low mobility of ions.

The drop in the likelihood of more elongated defects towards higher temperatures up to the critical temperature is a consequence of changes in the line and surface tensions, but also the films' ability to restructure. 
As temperatures allow the film to turn more liquid, high ion mobility promotes self-curing of defects, so more elongated defects either contract to a nearly circular shape and/or collapse more frequently. 
Actually, the results show that above the critical temperature, there is a characteristic size ($A$) and shape (IPR) of a defect, with the latter being slightly higher than at $T_c$ due to increased fluctuations. 
Given link between hole size and uncovered surface area it is likely that the typical size depends on the overall surface coverage.

On the other hand, the characteristic shape, unlike the average number and size of holes, should not be directly tied to the coverage.
Its impact should consequently be limited to controlling the theoretically possible maximum and minimum of these distributions.
On close inspection, at \SI{350}{\kelvin} and \SI{400}{\kelvin} at this low degree of surface saturation, we observe the distributions of $P_\mathrm{hole}$ and $A_\mathrm{hole}$ shifting towards larger values with temperature, while the number of holes doubles and even quadruples compared to \SI{300}{\kelvin}.
As a consequence of these larger holes with a longer outline, we see the lower limit of the IPR distribution moving closer towards the theoretical minimum associated with a circle.
Higher surface saturation is expected to limit the ability of holes to better approach circular shapes as well as the formation of more elongated shapes at the same.
However, thermal fluctuations are always expected to cause the spread of the morphological distributions to rise along with $T$ above $T_c$ (\cref{fig:hole_morphology}) as far as the limitations set through surface coverage allow.

\subsection{Morphology of the film}
\begin{figure*}
    \centering
    \includegraphics[width=.33\textwidth]{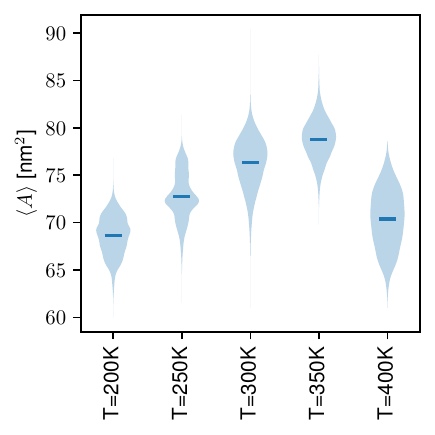}%
    \includegraphics[width=.33\textwidth]{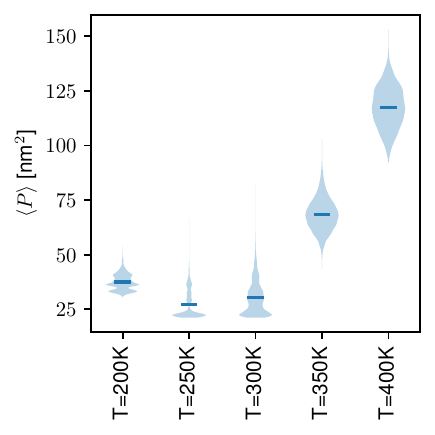}
    \includegraphics[width=.33\textwidth]{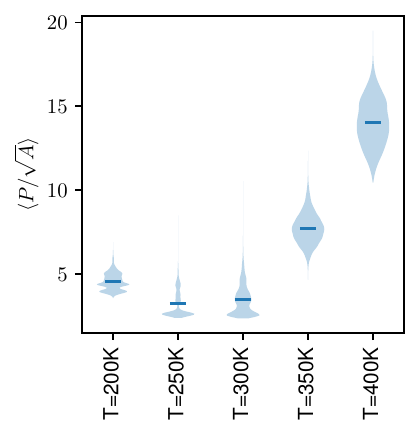}%
    \caption{\textbf{Statistical analysis of the monolayer film morphology at various temperatures.}
    (left panel) The mean surface area covered per 100 liquid film molecules increases slightly with temperature. 
    The normalized perimeter (middle panel), however, grows more significantly in the liquid state due to the dynamic formation of defects. 
    This leads to an overall increase in the isoperimetric ratio $P/\sqrt{A}$ (right panel).
    (Detailed data about the statistics visualized here provided on ZENODO \cite{zenodo_data_archive})}
    \label{fig:film_morphology}
\end{figure*}

Further analysis can be performed based on the morphology of the overall film. 
This analysis involves calculating the mean area $\langle A\rangle$ covered by the monolayer film and determining its mean outline length/circumference $\langle P\rangle$ as a function of the temperature (see \cref{fig:film_morphology}a and b). 
While we generally observe a trend of increased surface area being covered per ion pair at larger temperatures, the relative change is not exceeding \SI{15}{\percent}.

Only at \SI{400}{\kelvin} do we observe a drop in the mean area per ion, which is a consequence of thin strings of ions lining the perimeter of holes forming at that temperature. 
For such structures, the current graph representation underestimates the covered area. 
Imposing a minimum thickness upon these strings would likely bring the results for \SI{400}{\kelvin} more in line with the general trend in which the ion distance increases as expected, due to thermal fluctuations at higher temperatures.

For the circumference of the film, the increase is, however, much more significant. 
The change is rather small below the critical temperature, indicating that the film is minimizing its perimeter under constraints imposed by the underlying substrate. 
A deviation from the trend of increasing perimeters with temperature is, therefore, only observed between \SI{200}{\kelvin} and \SI{250}{\kelvin}, which can be attributed to the slow dynamics at very low temperatures, and the inability of the film to find a fully relaxed configuration on the time scales of the simulations. 
This then again agrees with the previously mentioned experimental findings where approaching the transition temperature from below is associated with  the IL monolayer film minimizing its circumference \cite{meusel2020growth}.

As $T$ exceeds $T_c$, the interaction of ions with the sapphire weakens, viscosity decreases, and ions get to explore a significant surface area, hence significantly increasing the entropy. 
The consequences are the more common formation of transient defects and longer ion strings, strongly increasing the circumference of the film.

Looking at the IPR of the film in \cref{fig:film_morphology}c, we see that changes in the perimeter dominate changes in the surface area, a finding emphasized by the  small size of our systems relative to the macroscopic films. Consequently, trends in IPR of the film as a function of temperature are identical to that of the perimeter. 
This agrees with our understanding that close to $T_c$, the film is able to balance the constraints from its coupling to the substrate with thermal excitations. 
Below $T_c$, the enthalpy of interaction is much larger than the  entropy. 
Restructuring is so slow that defects practically freeze and an optimal structure on the level of the film may not be fully attained. 

At higher $T>T_c$, entropy dominates enthalpy, the hole formation is frequent and IPR is increased.
It is, however, still notable, that while we observe the formation of thin strings of ions as well as of fingers pointing into the uncovered substrate surface area, detached colonies like the one visible in \cref{fig:system_illustration}b and c are usually small and, furthermore, rare occurrences.
This suggests that even at the larger simulated temperatures, it is still energetically favorable for the film to remain connected.
At even larger temperatures, we expect the thermal excitation to also overcome this constraint leading the film to decompose into independent groups of ions.
However, such effects could occur at temperature that are beyond those at which the ions themselves become unstable and decompose  \cite{heym2011analysis}.

\section{The effect of confinement on the organization of IL monolayers}

An important observation from the previous analysis is that the monolayer remains intact at all temperatures and no spontaneous formation of bilayers or multilayers has been observed. 
This is notable, as the bi-layer formation is an effect of the competition between the out-of-plane adhesive forces with the substrate and in- and out-of-plane cohesive forces within the ionic liquid.
Our results suggest that the strong adhesion forces in our system are able to outweigh the out-of-plane cohesive forces. 
Furthermore, the in-plane cohesive forces are so strong that up to the highest temperature in our simulations it was impossible to just pull one ion out of the film without dragging a paired ion when calculating the PMF.
However, an increase in the saturation of the interface layer has been shown to lead to a change in the observed behavior \cite{meusel2020growth}. 
Actually, at low temperatures, the experiments observed islands of the IL on the gold surface with the second layer present on top of the wetting layer, forming out-of-plane checkerboard patterns. Furthermore, for a related IL with chloride anions between graphite surfaces, the confinement has been observed to play a major role in structuring \cite{sha2008liquid}.
It specifically caused a phase transition from a liquid (isotropic) to a solid (directional) structure upon stronger confinement. 
One can therefore argue that the stability of our monolayer is a consequence of the surface not being fully covered by the film.

In our systems, we presume the bilayer structures to be metastable, relative to the monolayer. 
The evidence for this effect can be found in observing the spreading dynamics (\cref{fig:wetting}). 
At all temperatures, following a quick expansion of the droplet, the precursor monolayer film forms. 
The spreading is continuous until the droplet is reduced to a bilayer above a part of the monolayer, when we typically observe an arrest in the dynamics. 
At some point, however, the pure monolayer is formed with a quick disappearance of the bilayer, which can be seen from the abrupt change of slope at low effective spreading angles. 
This is reminiscent of transitioning from a metastable into a more stable configuration, over a barrier. 
At low temperatures, crossing over this barrier is very difficult as the mobility of ions after they form hydrogen bonds is very low, and hence the spreading is overall impeded. 

\begin{figure*}
    \centering
    \includegraphics[width=1\textwidth]{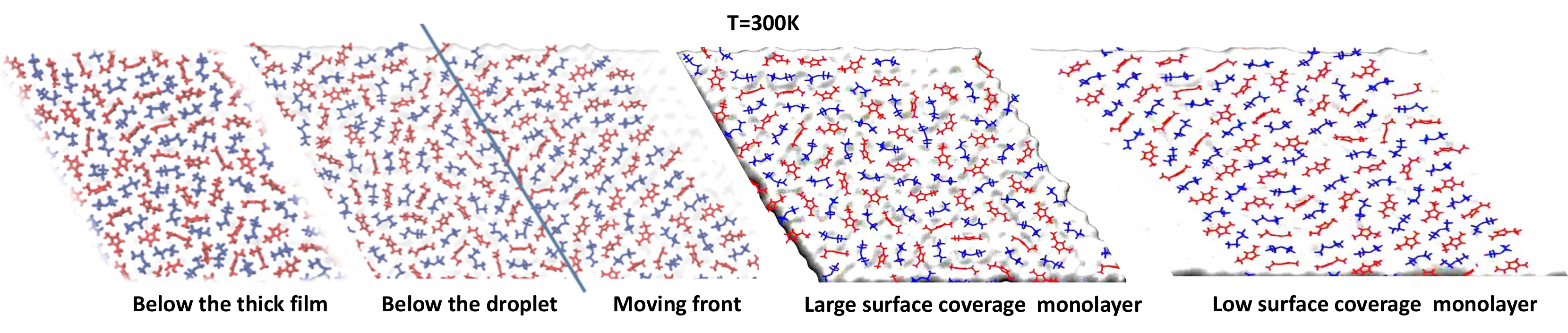}
    \caption{\textbf{Effect of confinement on the structure of the film close to the monolayer transition temperature of $T=\SI{300}{\kelvin}$.}
    Short range order is maintained below a thick film, as well as under the droplet and a monolayer at relatively large surface coverage. 
    Long range order is, however, observed in the precursor film during spreading and even more clearly in an unconstrained monolayer, where only the part of the surface with the IL is shown.
    }
    \label{fig:frozen_layer_300K}
\end{figure*}

We therefore focus our discussion on the stability of the monolayer close to the critical temperature of \SI{300}{\kelvin}, when the bulk IL is clearly in the liquid state. 
At this point even below the thick film, the contact layer experiences very slow dynamics, but only short range order. 
The same type of (dis)order is found in the contact layer below the droplet but notably, also when we create the monolayer from a thick film, by removing all ions that are not in contact with the surface. 
Providing a larger area of the sapphire at this point produces long-range order. 
Yet under constraint of higher density, the system can either maintain long range order in the contact layer with a bilayer on top, hence gaining out-of plane cohesion energy, or sacrifice the order and stay in the monolayer configuration, which seems to be the case for our substrate-liquid combination. 
However, fully resolving the energetics of these states may require a more in-depth study.

These results demonstrate the importance of the balance between (i) adhesive forces and the specificity of interaction of the IL with the functional moieties in the support, 
(ii) the film-internal cohesive forces, causing alternating patterns of anions and cations, 
and (iii) out-of-plane cohesive forces between the different layers, which weaken the relative impact of the adhesive interactions with the interface, especially when the surface has already been fully covered. 
We therefore suggest that interface-adjacent behavior relating to adsorption in liquid films and diffusive transport may strongly depend on the overall film thickness. 
More importantly, this highlights the impact of the level of confinement on the position of the critical point in supported liquid films. 

\section{Discussion and Conclusions}

In our analysis of the structural and morphological properties of monolayer films of [C$_2$Mim][NTF$_2$] on a neutral, hydroxylated alumina substrate, we observe two clear regimes for the behavior of liquid films.
The strong coupling between the nearly immobile ions and the underlying crystal structure dominates at low temperature. 
This results in a very distinct directionality and periodicity of the films' internal structure, not usually observed in a liquid phase.
We attribute this to adhesive forces and an effective confinement of the liquid layer through the formation of hydrogen bonds, similar to those observed in density functional calculations and other simulations. \cite{sha2008liquid,zhang2018structure}
In contrast, at higher temperatures, thermal excitations dominate, and ions start to hop between surface binding sites, while still being adsorbed. 
This decouples the internal film structure from the structure of the underlying alumina. 
In this effectively less confined regime, the alternating order of ions of opposing charge is retained, dominating the internal film  organization. 
It is the optimization of cohesive in-plane Coulomb-interactions that dominates the patterning process. 
Consequently, the directionality imprinted on the monolayer by the substrate is lost. 

To quantify the extent of the changes associated with the observed structural phase transition, we analyzed the morphology of the monolayer films and the defects appearing spontaneously therein.
Below the critical temperature, the defects are rare, nearly frozen, little fluctuating in size. 
On-average, however, the lowest probability for observing a hole is at around $T_c$, as at this temperature, the mobility of ions is just enough to heal an opening in the monolayer. 
The film is minimizing its circumference and its isoperimetric ratio, which is reflected in the minimum in the isoperimetric ratio of the holes.

Above the critical temperature, the area and perimeter of individual holes remains approximately constant, while there is a clear increase in the frequency of their occurrence. 
At this point, the IPR of the film increases also due to the formation of chains of ions alternating in type.
These strings increase the circumference of the film without significantly contributing to the covered area, and are a consequence of thermally reduced line tension.

Our data show that the precise morphology of the monolayer depends on the degree of saturation of the ionic layer on top of the substrate, in accordance with previous experimental findings \cite{meusel2020growth}.
Specifically, as surface coverage increases, the total area of all holes is constrained. 
This forces defects to attain smaller sizes, making them increasingly indistinguishable from the remainder of the monolayer, at the expense of the enthalpy associated with in-plane Coulomb forces. 
Our previous analysis of the contact layer of the IL with a sapphire support below a several nanometers thick film showed that the mobility of ions is basically impeded, but the structure is not fully ordered, even at room temperature \cite{epm_paper,vuvcemilovic2021computational}.
We, thus, expect an increase in out-of-plane ion pair interactions competing with surface attraction to cause an even earlier loss of order.
The precise threshold temperature $T_c$ is, therefore, expected to shift to lower values due to differences in system geometry, saturation of the interface layer and film thickness. 
These changes tie back to a loss of confinement, which has previously been linked to structural properties of thin films \cite{sha2008liquid,zhang2018structure}.
However, more quantitative analysis of these effects is warranted.
Future work should, therefore, focus on the formation of bi-layers or thin multi-layers to find the relation between transitional behavior in highly confined mono- and low-confinement multi-layer film systems.

\printcredits

\section{Acknowledgements}
The authors declare no competing financial interest. We thank G. Hantal for a critical proof read of the manuscript. We acknowledge funding by the Deutsche Forschungsgemeinschaft (DFG, German Research Foundation) – Project-ID 416229255 – SFB 1411 Particle Design and Project-ID 431791331 - SFB 1452 Catalysis at Liquid Interfaces (CLINT) for support.
The authors gratefully acknowledge the scientific support and HPC resources provided by the Erlangen National High Performance Computing Center (NHR@FAU) of the Friedrich-Alexander-Universit\"{a}t Erlangen-N\"{u}rnberg (FAU).



\FloatBarrier

\bibliographystyle{model1a-num-names}


\bibliography{bibliography}



\end{document}